\documentclass[12pt]{article}
\usepackage{subfigure,a4}
\usepackage{multirow}
\usepackage{epsfig,ulem}
\usepackage{dcolumn}% Align table columns on decimal point
\usepackage{bm}% bold math

\newcommand{\be}{\begin{eqnarray}}
\newcommand{\ee}{\end{eqnarray}}

\usepackage[english]{babel}
\usepackage{amssymb}
\usepackage{amsmath}
\usepackage{amsfonts}
\usepackage{amsbsy}
\usepackage{indentfirst}
\usepackage{graphicx}
\usepackage{color}

\newcommand{\bfgr}{\boldsymbol{\nabla}}
\newcommand{\bfal}{\boldsymbol{\alpha}}

\def\d{{\rm d}}

\begin{document}

%%%%%%%%%%%%%%%%%%%%%%%%%%%%

%\title{Ferromagnetic condensation in high density hadronic matter}

\title{Ferromagnetic condensation in high density hadronic matter}
\author{
Henrik Bohr\\
{\it \small Department of Physics, B.307, Danish Technical
University,}\\{ \it \small DK-2800 Lyngby, Denmark}\\Prafulla K.
Panda\\{\it\small Departmenet of Physics, C.V. Raman College of
Engineering,}\\{ \it \small  Vidya Nagar, Bhubaneswar-752054, India, and}\\
{\it \small CFC, Departamento de F\'\i sica, Universidade de Coimbra,}\\
{\it \small  P-3004-516 Coimbra, Portugal} \\
Constan\c{c}a Provid\^encia,
Jo\~ao da Provid\^encia\\
{\it \small CFC, Departamento de F\'\i sica, Universidade de Coimbra,}\\
{\it \small  P-3004-516 Coimbra, Portugal} }
% \maketitle

\maketitle
\begin{abstract}
We investigate the occurrence of a ferromagnetic phase transition in
high density hadronic matter (e.g., in the interior of a neutron
star). This could be induced by a four fermion interaction analogous
to the one which is responsible for chiral symmetry breaking in the
Nambu-Jona-Lasinio model, to which it is related through a Fierz
transformation.  Flavor $SU(2)$ and flavor $SU(3)$ quark matter are
considered.
%In flavor $SU(2)$ quark matter, a
A second order phase transition is predicted at densities about 5
times the normal nuclear matter density, a magnetization of the
order of $10^{16}$ gauss being expected. It is also found that in
flavor $SU(3)$ quark matter,  a first order transition from the
so-called 2 flavor super-conducting (2SC) phase to the ferromagnetic
phase arises.
%The
%ferromagnetic phase is more stable for $\mu>0.51$ GeV. The 2SC phase
%is more stable for $\mu<0.51$ GeV.
The color-flavor-locked (CFL)
phase may be completely hidden by the ferromagnetic phase.

\end{abstract}

\section{Introduction}
%Introduction
The %magnetic
properties of a highly compressed fermion fluid, especially with
respect to magnetization, are of great interest since they are
relevant for stellar objects such as neutron stars and, possibly,
quark stars \cite{haensel}. Recently, magnetars, which are a kind of
neutron stars exhibiting extremely powerful magnetic fields, have
been discovered \cite{g,goegues}.

We investigate a model of the magnetization of a quark fluid which
is characterized by a four fermion interaction \cite{bohr}  related
to the interaction of the standard Nambu-Jona-Lasino (NJL) model
through a Fierz transformation. In the present model, a
ferromagnetic transition is induced by an analogous mechanism to the
one which is responsible for chiral symmetry breaking in the NJL
model \cite{njl,klevansky,hatsuda,buballa}. This model
%which is now under investigation
accounts for the ferromagnetic transition at very high densities,
after chiral symmetry restoration. This is at variance with
ref.\cite{broniowski}, where it is shown, in the framework of the
chiral model of pions and quarks, also known as the linear sigma
model, that spin polarization of hadronic matter may arise as a
consequence of pion condensation, if chiral symmetry has not yet
been restored.

It should be pointed out that, in terms of a Skyrme force, a
ferromagnetic transition has been obtained for  nuclear matter
\cite{navarro,skyrme,vidana}. However, in that case, microscopic
calculations do not predict a ferromagnetic transition, at least
below $\rho=7\rho_0$ \cite{isaac02}. A ferromagnetic phase of quark
matter, has recently been investigated by several authors
\cite{tatsumi,iwazaki,ebert}, who focus on its QCD origin.

\section{Symmetrical quark matter}

%We are concerned with  so
The model we are considering applies to high  densities where chiral
symmetry has been restored. Hence, the
%in the sense that
quark masses are reduced to their bare masses, which are assumed to
vanish. Our model is of the NJL type
\cite{njl,klevansky,hatsuda,buballa}, but, for simplicity, we
neglect the typical scalar-scalar and pseudoscalar-pseudoscalar
interaction which is responsible for chiral symmetry breaking.  This
is because, at very high densities, it has no effect. Instead, we
construct a pure fermionic QCD inspired model based on a chiral
symmetric \color{black} tensor-tensor interaction, so the Lagrangian
reads
$${\cal
L}=i\bar\psi\gamma^\mu\partial_\mu\psi-{1\over4}G[(\bar\psi\gamma^\mu\gamma^\nu\tau_k\psi)(\bar\psi\gamma_\mu\gamma_\nu\tau_k\psi)
-(\bar\psi\gamma^\mu\gamma^\nu\gamma_5\psi)(\bar\psi\gamma_\mu\gamma_\nu\gamma_5\psi)].$$
We stress that this interaction may be understood as arising from a
the Fierz transformation of a standard NJL interaction.  We consider
$\mu,\nu$ of the form $i,j\in\{1,2,3\},~i\neq j.$ In the following
we will consider massless quarks and we will work in the no-sea and
mean-field approximations. By mean-field approximation we mean that
the operator $\bar\psi\gamma_i\gamma_j\tau_k\psi$ is replaced by
$\langle\bar\psi\gamma_i\gamma_j\tau_k\psi\rangle$
$+\bar\psi\gamma_i\gamma_j\tau_k\psi-\langle\bar\psi\gamma_i\gamma_j\tau_k\psi\rangle$
and  second order terms (and  higher) in
$\bar\psi\gamma_i\gamma_j\tau_k\psi-\langle\bar\psi\gamma_i\gamma_j\tau_k\psi\rangle$
are neglect.  In the Conclusions, the implications of the no-sea
approximation will be briefly discussed.  Let us assume polarization
along the 3 axis. Then, $(\bar\psi\gamma_1\gamma_2\tau_k\psi)^2$ is
replaced by
$2\langle\bar\psi\gamma_1\gamma_2\tau_k\psi\rangle\bar\psi
\gamma_1\gamma_2\tau_k\psi-\langle\bar\psi\gamma_1\gamma_2\tau_k\psi\rangle^2.$
We introduce the notation
\begin{equation}
F_k=G\langle\bar\psi\Sigma_3\tau_k\psi\rangle=iG\langle\bar\psi\gamma_1\gamma_2\tau_k\psi\rangle.
\label{fk}
\end{equation}
Manifestation of this interaction is seen only at very high
densities. In the mean field approximation, the postulated
interaction leads to the Lagrangian {density}\begin{equation}{\cal
L}_{MFA}=i(\overline\psi\gamma^\mu\partial_\mu
\psi)-F_k(\overline\psi \Sigma_3\tau_k\psi)-{F_k^2\over
2G}.\end{equation} In turn, this leads to  the Dirac equation
%$$(-i
%\partial_t+i\alpha_j\nabla_j)
%\psi-F_k \beta \Sigma_3\tau_k\psi=0,$$
%which is equivalent to
$$(-i\bfal\cdot{\bfgr}+\epsilon_\tau
F\beta\Sigma_3)\psi=\varepsilon\psi,$$ where $\epsilon_\tau=1$ for
quarks $u$  and $\epsilon_\tau=-1$ for quarks $d$ denote the
eigenvalues of $\tau_3$, and  $F_k$ is a parameter which is
determined  from (\ref{fk}).  In the following, we take $F_1=F_2=0,$
and $F_3=F$.

In this section we will consider symmetric 2-flavor quark matter,
the Fermi energy being the same for quarks $u$ and $d$.
%with $\mu_u=\mu_d$.
The single particle energy eigenvalues satisfy
\begin{equation}\varepsilon_p=\pm\sqrt{\left(|F|\pm\sqrt{p_1^2+p_2^2}\right)^2+p_3^2}~,\label{epsilon}
\end{equation}
where the $\pm$ signs are such that the single particle energies of
quarks $u$ with momentum $\bf p$ and spin $s$ are the same as the
energies of quarks $d$ with the same momentum  and spin $-s$. By
$p_3$ we denote the component of momentum along the direction of
spin polarization, $p_1$ and $p_2$ being the transverse components.
A {\it no sea} approximation shall be used. Thus, the quark Fermi
surfaces are given by
$$\mu^2={\left(|F|\pm\sqrt{p_1^2+p_2^2}\right)^2+p_3^2}~,$$
 the $+$ sign corresponding to the Fermi surface of quarks $u$ with
spin up or quarks $d$ with spin down, (in the direction of
magnetization) and the $-$ sign to the Fermi surface of quarks $u$
with spin down or quarks $d$ with spin up. Since the temperature is
assumed to vanish, the Fermi energy coincides with the chemical
potential $\mu$.

\begin{figure}[ht]
\centering
\includegraphics[width=0.8\textwidth, height=0.5\textwidth]
{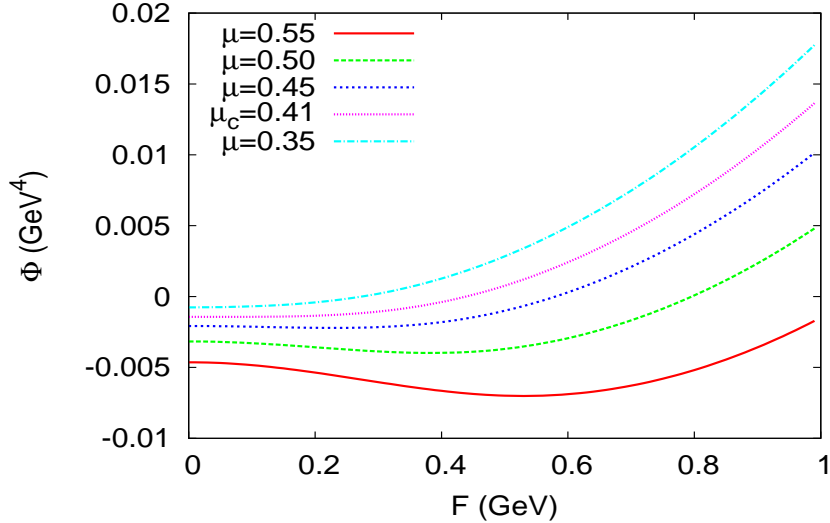}
\caption{The thermodynamical potential $\Phi$
vs. $F$, for symmetric flavor $SU(2)$ quark matter, for fixed values
of $\mu$, being $\mu_c=0.41$ GeV.
%{\color{red}Replace the curve for mu=0.4 by a curve for mu=0.41}
} \label{fig0}
\end{figure}

%%%%%%%%%%%%%%%%%
We observe that the formalism behind the description of
ferromagnetic condensation is analogous to the BCS theory (the
parameter $F$ playing the role of the gap $\Delta$), or to the NJL
theory for chiral symmetry breaking  (the order parameter $F/G$
corresponding then to the quark condensate $\langle\bar q
q\rangle$). The condition
\begin{equation}\label{gap}{\partial\Phi\over\partial
F}=0\end{equation} replaces the gap equation. Below the critical
potential $\mu_c$, this condition implies that $F=0$, so that the
state of equilibrium is the normal phase. Above $\mu_c$, equilibrium
occurs for $F\neq0$ (cf. Fig \ref{fig0}). The critical chemical
potential $\mu_c$ is determined by the condition
$$\left.{\partial^2\Phi(\mu,F)\over\partial F^2}\right|_{F=0}=0$$
and is given by $$\mu_c={\pi\over\sqrt{3G}}.$$

The Fermi surface may be visualized as a circle of radius $\mu$
rotating around a line lying on its plane at the distance $F$ from
its center. If the line lies outside the circle, the Fermi surface
is doughnut-shaped and we have full polarization. If the line
crosses the circle but is not a diameter, we have two Fermi
surfaces, one for spin up and the other one for spin down,
corresponding to partial polarization. If the line is a diameter, we
have no polarization at all, since the Fermi surface is the same for
particles with spin up or down. Thus, two different expressions are
obtained for the thermodynamical potential $\Phi(F,\mu)$, one which
is valid for $|F|>\mu$ (needed at higher densities), and another one
for $|F|<\mu$ (needed at lower densities).

Taking into account that the color-flavor degeneracy is 6, and
keeping in mind that $2\pi^2\mu^2F$ is the volume of the torus
generated by a circle of radius $\mu$ whose center describes a
circle of radius $F$, it is easily found that, for $|F|>\mu$, when
full polarization is present, the particle number reads,
\begin{equation}
N={{6}V\over(2\pi)^3}\int\d^3{\bf p}\theta(\varepsilon_p-\mu)={12V\over(2\pi)^3}\pi^2
|F|\mu^2~,
\label{n}
\end{equation}
\begin{figure}[ht]
\centering
\includegraphics[width=0.6\textwidth, height=0.5\textwidth]{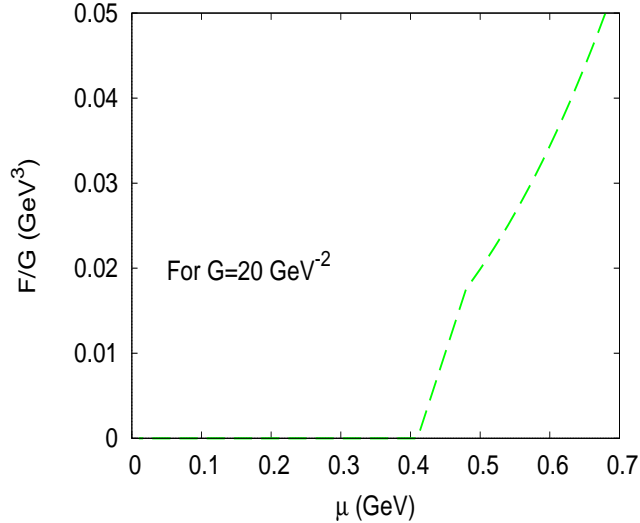} 
\caption{The order parameter $F/G $ versus the
chemical potential $\mu$, for symmetrical matter (SM). In the normal
phase, $F/G$=0.
%{\color{red}Units of F/G (GeV3)}
} \label{fig1}
\end{figure}
 and the energy reads,
\begin{equation}
E={{6}V\over(2\pi)^3}\int\d^3{\bf
p}\theta(\varepsilon_p-\mu)\epsilon_p+{VF^2\over2G}={8V\over(2\pi)^3}\pi^2
|F|\mu^3+{VF^2\over2G}~,
\label{ener}
\end{equation}
where the term ${VF^2/(2G)}$ comes from the standard mean field
approximation. The limits of integration over the transverse
momentum $\sqrt{p_1^2+p_2^2}$ in Eqs. (\ref{n}) and (\ref{ener})
%depends on the value of
are $F-\mu$ and $F+\mu$.
%The integrals in (\ref{n}) and (\ref{ener}) are
%easily obtained.
The thermodynamical potential becomes
$$\Phi=E-N\mu=-{4V\over(2\pi)^3}\pi^2
|F|\mu^3+{VF^2\over2G}~.$$ This is the expression which is valid for
$F\geq\mu$. The physical $F$ value minimizes this expression and is
equal to
$$F={4}G{\pi^2\mu^3\over(2\pi)^3}~,$$so that
in the fully polarized  phase the thermodynamical potential reads
$$\Phi=-{V\over\pi^2}~{G\mu^6\over8}.$$
However, this expression is valid only  for
$\mu\geq\mu_p=\sqrt{2\pi/G}$, since $\sqrt{2\pi/G}$ is the $\mu$
value such that minimum of $\Phi$ occurs precisely for $F=\mu$. For
$\mu<\mu_c$, the condition (\ref{gap}) implies that $F=0$, so that
the state of equilibrium is the normal phase  for which the
thermodynamical potential equals
$$\Phi=-{V\over\pi^2}~{\mu^4\over2}.$$
\begin{figure}[ht]
\centering
\includegraphics[width=0.6\textwidth, height=0.5\textwidth]{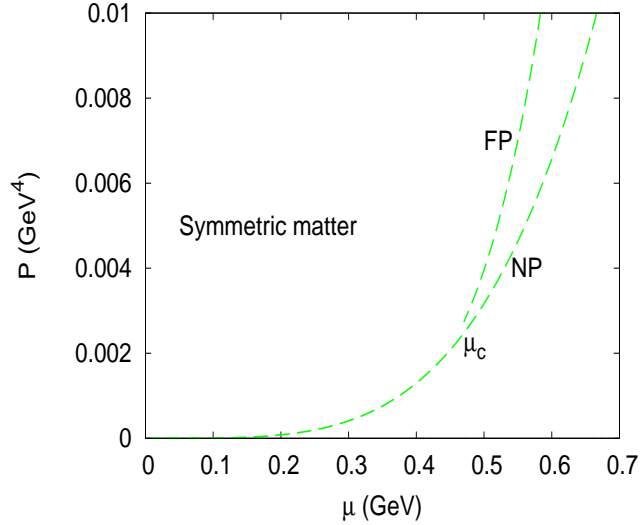} 
\caption{The pressure $P$ vs. the chemical
potential $\mu$ for $G=20$ GeV$^{-2}$ comparing the ferromagnetic
phase (FP) and the normal phase (NP) of symmetric matter.
%Only values of $\mu$ up to 0.5 are relevant.
} \label{fig2}
\end{figure}

In order to describe the system for $\mu_c<\mu<\mu_p$, we need
$\Phi(\mu,F)$ for $0<F<\mu.$  This is  the case of partial
polarization, which we now  briefly discuss.
% that is $\mu>F>0$.
Then, the limits of integration over the transverse momentum
$\sqrt{p_1^2+p_2^2}$ are $0,\mu-F$ (spin up) and $0,\mu+F$ (spin
down), so that the quark number equals,
\begin{eqnarray}&&\nonumber N(\mu,F)={6V\over(2\pi)^3}\left(\int_\uparrow\d^3{\bf
p}\theta(\varepsilon_p-\mu)+\int_\downarrow\d^3{\bf
p}\theta(\varepsilon_p-\mu)\right)\\&&={V\over\pi^2}\left(\sqrt{\mu^2-F^2}
(F^2+2\mu^2)+3F\mu^2\tan^{-1}{F\over{\sqrt{\mu^2-F^2}}}\right)\label{N(mu,F)}.\end{eqnarray}
For the energy we find
$$E=E_1+{VF^2\over2G}$$\begin{figure}[ht]
\centering
\includegraphics[width=0.6\textwidth, height=0.4\textwidth]{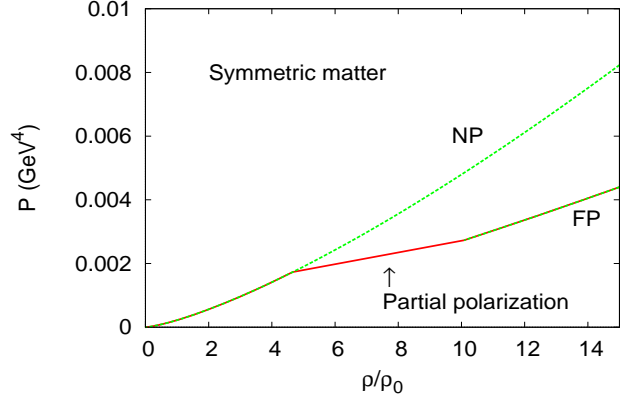} 
\caption{The pressure $P$ vs. the quark density
$\rho$ for $G=20$ GeV$^{-2}$, for symmetrical quark matter, in the
ferromagnetic (FP) and in the normal (NP) phases. Partial
polarization is described by the portion of the graph with lowest
slope, which occurs for $ 4.87\rho_0\leq\rho\leq~ 9.99\rho_0~$ and
for $0.00175<P<0.00273~({\rm GeV}^4).$
 } \label{fig3}
\end{figure} where
\begin{eqnarray}&&\nonumber E_1(\mu,F)={6V\over(2\pi)^3}\left(\int_\uparrow\d^3{\bf
p}~\varepsilon_p~\theta(\varepsilon_p-\mu)+\int_\downarrow\d^3{\bf
p}~\varepsilon_p~\theta(\varepsilon_p-\mu)\right)\\&&\nonumber=\int^\mu_{0}
\d\mu~\mu{\partial N\over\partial\mu}=\mu N-\int^\mu_{0} \d\mu~ N
\\&&=\mu N-\int^F_{0} \d\mu~ N-\int^\mu_{F} \d\mu~ N=\mu
N-{4V\over8\pi^2}F^4+\Phi_1,\label{E1(mu,F)}
\end{eqnarray}
\color{black} In the case $0\leq|F|\leq\mu$, the thermodynamical
potential $\Phi$ is given by
\begin{eqnarray}\Phi(\mu,F)=\Phi_1-{4V\over8\pi}F^4+{VF^2\over2G},\label{Phi}\end{eqnarray}where,
%\color{red}
\begin{eqnarray}\label{Phi1}&&\Phi_1(\mu,F)=-\int^\mu_{F} \d\mu~ N
= -{VF^4\over\pi^2} \left({2\mu^3+3\mu F^2\over4
F^4}\sqrt{\mu^2-F^2} \right.\nonumber\\&&\left.+{\mu^3\over
|F|^3}\tan^{-1}{|F|\over\sqrt{\mu^2-F^2}}
-{1\over4}\log{\mu+\sqrt{\mu^2-F^2}\over
|F|}-{\pi\over2}{}\right).\end{eqnarray}
%Please check.
In (\ref{Phi}), $|F|$ is determined by minimizing $\Phi$ %with respect to $F$
for fixed $\mu$. It is found that for $\mu<\mu_c$, the minimum of
$\Phi(\mu,F)$ occurs for $F=0$, so that, for low densities, the
normal phase prevails. For $\mu_c<\mu<\sqrt{2 \pi/G}=\mu_p$ the
minimum of $\Phi(\mu,F)$ occurs for $0\leq|F|\leq\mu$, and then
partial polarization is realized.

%%%%%%%%%%%%%%%%%%%%%%%%%%%%%%%%%%%%%%%%%%%%%%%%%%%%%%%%%%%%%%%%%
 Numerical results for  $G=20$ GeV$^{-2}$,
arbitrarily chosen but of the order of the coupling constant of  the
chirally symmetric four-fermion interaction of the NJL model which
takes values $\sim 5-10$ GeV$^{-2}$
%\cite{bub}
\cite{buballa}, are
presented in Figs. \ref{fig1} to \ref{fig3}.

In Fig. \ref{fig1}, the order parameter $F/G $ is plotted versus the
chemical potential $\mu$, for symmetrical matter. The ferromagnetic
phase occurs for a quark chemical potential  above $\mu_c=$0.41 GeV.
A transition of second order, characterized by  a continuous
behavior of the magnetization and of the baryonic density, is found.
%The phase transition was determined using Gibbs criterium, namely,
%at the transition, the pressure and the chemical potential are the
%same in both phases.

In Fig. \ref{fig2}, the pressure $P$ is plotted versus the chemical
potential $\mu$ for both the ferromagnetic and the normal phases of
symmetric quark matter. The most stable phase is the one having  the
largest pressure for a given chemical potential. Therefore, for the
lowest chemical potentials the normal phase is the equilibrium phase
for both symmetric and neutral matter,
%(full lines in Fig.\ref{fig2}),
while at high chemical potential the ferromagnetic
phase
%(dashed lines)
is the stable one.
\begin{center}
\begin{figure}[!htb]
\subfigure[ The critical chemical potentials $\mu_c$ and $\mu_p$.]
{\includegraphics[scale=0.55]{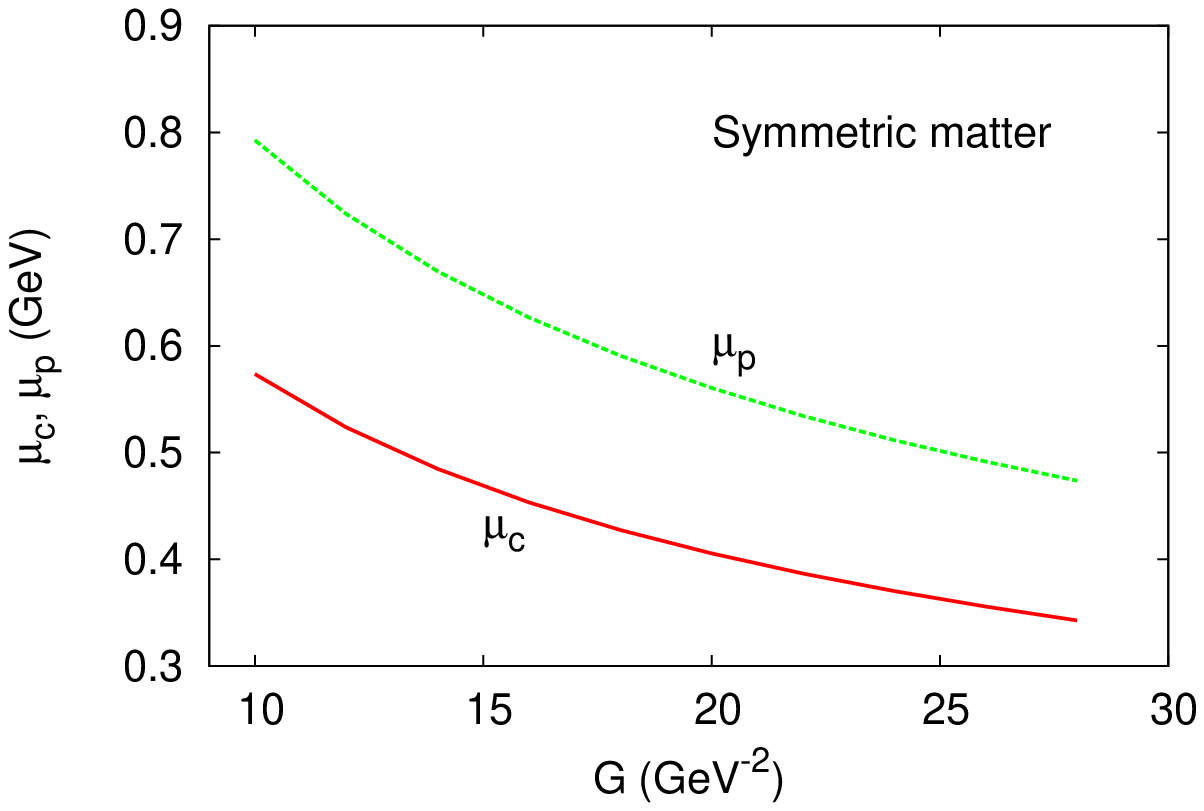}} 
\subfigure[The critical pressure $P_c$ ]
{\includegraphics[scale=0.55]{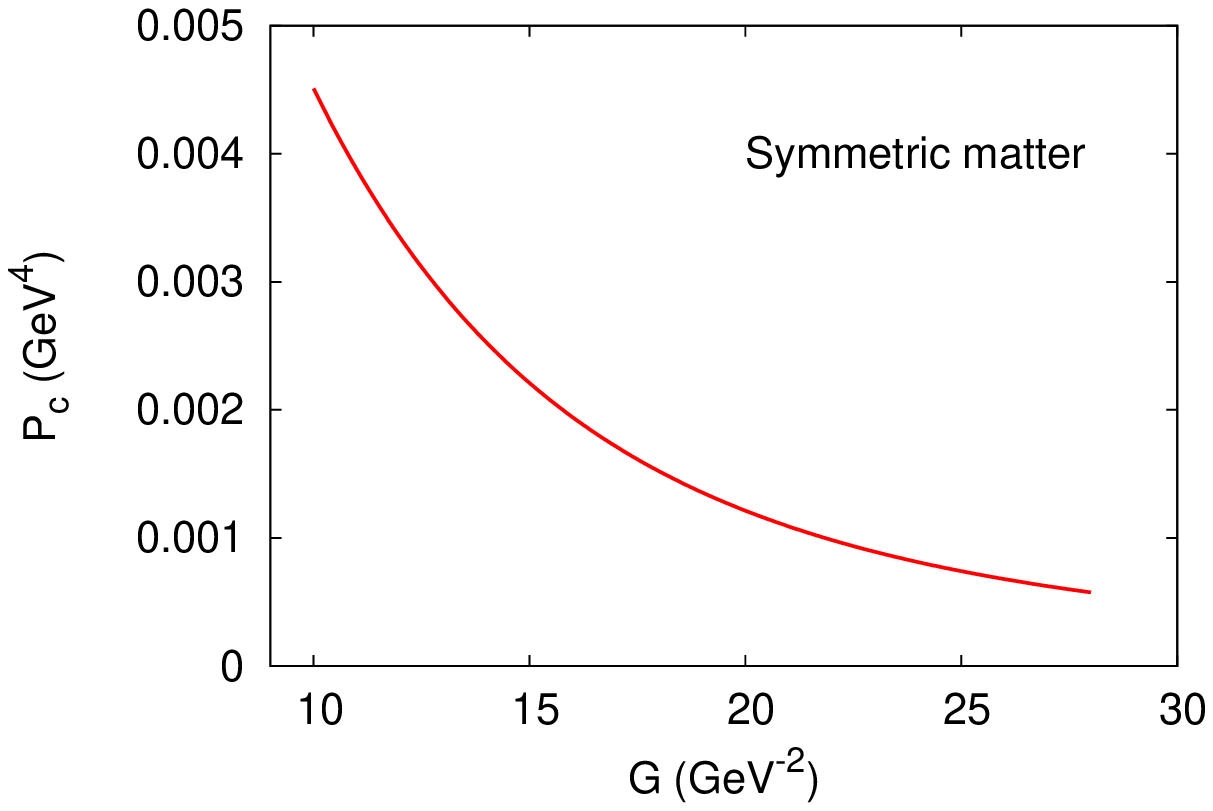}} 
\caption{The critical values of the pressure and of the chemical potentials
versus the coupling constant $G$. By $\mu_p $ we denote the critical
chemical potential for the onset of full polarization.} \label{fig4}
\vspace*{-0.2cm}
\end{figure}
\end{center}
 In Fig. \ref{fig3}, the pressure
$P$ is plotted versus  $\rho/\rho_0$, where $\rho$ is the baryonic
density and $\rho_0$ is the nuclear saturation density, comparing
the ferromagnetic and the normal phases of symmetrical quark matter.
Partial polarization is described by the portion of the graph with
lowest slope.  For symmetrical matter, this region  occurs for $
4.87\rho_0\leq\rho\leq~ 9.99\rho_0$,   and $0.00175<P<0.00273~({\rm
GeV}^4).$ The critical quark densities are quite high. A larger
coupling constant would induce a transition at lower densities. In
Fig. \ref{fig4}, the dependency of the critical values of the
pressure and of the chemical potential on the coupling constant $G$
is displayed. A phase transition at densities comparable to the ones
expected to exist in the interior of neutron stars,  twice to four
times the equilibrium density of nuclear matter, would require that
a $G$ of the order $25$ GeV$^{-2}$ to $30$ GeV$^{-2}$. %However,
Although these couplings are probably too large, they may be
justified, as explained below, if the vacuum polarization is taken
into account. Only values of $\rho$ up to $ 10\rho_0$ are relevant.

\section{Magnetization}
\begin{figure}[ht]
\centering
\includegraphics[width=0.6\textwidth, height=0.5\textwidth]{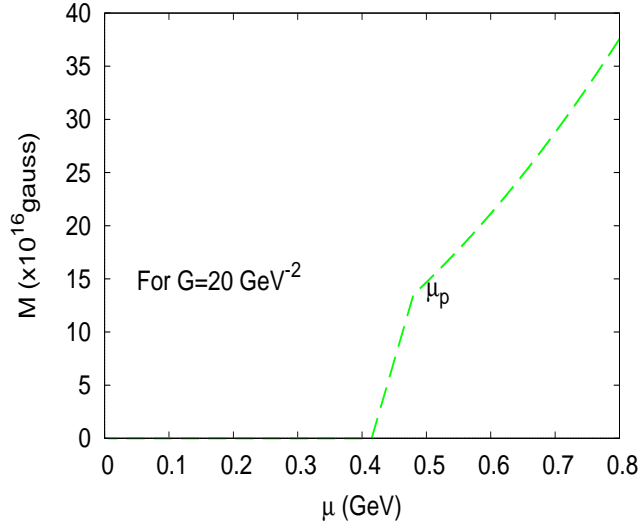} 
\caption{ Magnetization versus the chemical
potential $\mu$, for symmetrical matter, $\mu_p$ being the critical
$\mu$ for the onset of full polarized matter. In the normal phase,
${\cal M}=0$. } \label{fig5}
\end{figure}
As suggested in \cite{bohr}, magnetization of a relativistic fermion
fluid may be associated with the lowest Landau level (LLL). The
energy of the Landau level $\nu,p_3$ is obtained through the
replacement $p_1^2+p_2^2\rightarrow{ 2\nu |Q| B}$ in the expression
(\ref{epsilon}) of the single particle energy, so that, in the
presence of a magnetic field $B$, the energy of a quark with charge
$Q$ is equal to
$$\varepsilon_{\nu}(p_3)=\sqrt{\left(F\pm\sqrt{2|Q|B\nu}\right)^2+p_3^2}.$$
In our model, the {\it effective} LLL does not arise for $\nu=0,$
but for $\nu=F^2/(2|Q|B)$, so that its energy is
$\varepsilon_{LLL}(p_3)= |p_3|.$ By analogy with \cite{bohr}, we
postulate, somewhat ad hoc, that the contribution of the effective
LLL to the thermodynamical potential reads $$-{\cal
M}B={(|Q_u|+|Q_d|)
B\over2\pi}\sum_{p_3}{1\over2}\left(\varepsilon_{LLL}(p_3)-\mu\right),$$
where $|Q_\tau|$ is the quark charge and $|Q_\tau| B/(2\pi)$ the
Landau degeneracy. The magnetization of fully polarized matter is
then equal to
$${\cal M}=
{3 \mu^2 (|Q_u|+|Q_d|)\over8\pi^2}= { 3\mu^2
({2\over3}|e|+{1\over3}|e|)\over8\pi^2},\quad \mu>\mu_p.$$ For
$\mu_c<\mu<\mu_p $ we will have,
$${\cal M}=
{3 \mu^2
(|Q_u|+|Q_d|)\over16\pi^2}\left(2\mu(\mu-\sqrt{\mu^2-F^2})+F^2\log{\mu-\sqrt{\mu^2-F^2}\over\mu+\sqrt{\mu^2-F^2}}\right)
.$$ In fig. \ref{fig5}, the magnetization is plotted vs. the
chemical potential, for symmetric quark matter.
 \color{black}
\section{Neutral quark matter}
In electrically neutral quark matter, the density $\rho_d$ of quarks
$d$ is twice the density $\rho_u$ of quarks $u$. In the normal phase
$|F|=0$ and the numbers of quarks $u$ and $d$ are given by
$$N_u
={V\over\pi^2}\mu_u^3, \quad N_d ={V\over\pi^2}\mu_d^3,$$where
$\mu_u,~\mu_d$ are the Fermi energies of quarks $u,~d$,
respectively. %In neutral matter
Since $\rho_d=2\rho_u,$ %so that
we have $\mu_d=2^{1\over3}\mu_u.$
The thermodynamical potential is written
$\Phi=K_u+K_d-\mu(N_u+N_d)$. Here $\mu$ is the chemical potential
which fixes the quark number independently of flavor and the kinetic
energies are equal to $K_u ={3V\over4\pi^2}\mu_u^4, $ and $ K_d
={3V\over4\pi^2}\mu_d^4$,
 so that
$$\Phi={3V\over4\pi^2}(\mu_u^4+\mu_d^4)-\mu
{V\over\pi^2}(\mu_u^3+\mu_d^3).$$ The values of $\mu_u,\mu_d$ are
not fixed a priori but are determined by minimization of $\Phi$
%with respect to
under the condition $\mu_d=2^{1\over3}\mu_u$, which leads to
$$\mu_u={3\over1+2^{4\over3}}\mu,\quad \mu_d={3\cdot 2^{1\over3}\over1+2^{4\over3}}\mu,$$ so that
\begin{equation}
\Phi=-{3V\over4\pi^2}\left({3\over1+2^{4\over3}}\right)^3\mu^4.\end{equation}

In the fully polarized ferromagnetic phase, $|F|>\mu_d$ (since
$\mu_d>\mu_u$) we have (cf. eq. (\ref{n}))
$$N_u={3V\over4\pi}|F|\mu_u^2,\quad N_d={3V\over4\pi}|F|\mu_d^2.$$ Therefore,
for electrically neutral matter we obtain $\mu_d=\sqrt{2}\mu_u,$ and
$$K_u={V\over2\pi}|F|\mu_u^3,\quad K_d={V\over2\pi}|F|\mu_d^3.$$
The thermodynamical potential reads
$$\Phi={V\over2\pi}|F|(\mu_u^3+\mu_d^3)-\mu
{3V\over4\pi}|F|(\mu_u^2+\mu_d^2)+{VF^2\over2G}.$$ Under
minimization with respect to $\mu_u,\mu_d$ subject to the condition
$\mu_d=\sqrt{2}\mu_u,$ we find
$$\mu_u={3\over1+2^{3\over2}}\mu$$
so that
\begin{equation}
\Phi=-{27V|F|\over4\pi(1+2\sqrt{2})^2}\mu^3+{VF^2\over2G}.
\end{equation}
Upon minimization of the previous expression with respect to $F$ we
obtain
$$\Phi=-{3^6VG\over2^5(1+2\sqrt{2})^4\pi^2}\mu^6.$$

In order to describe the regime of partial polarization, the
condition that for each quark $u$ there are two quarks $d$ is
imposed by requiring that
\begin{equation}{1\over2}N(\mu_u,F)=N(\mu_d,F),\label{constraint}\end{equation}
where the function $N(\mu,F)$ is given by (\ref{N(mu,F)}). The
thermodynamical potential of neutral quark matter equals
\begin{equation}\Phi(\mu,F)={1\over2}\left(E_1(\mu_u,F)+E_1(\mu_d,F)\right)+{VF^2\over2G}-\mu\left({1\over2}
\left(N(\mu_u,F)+N(\mu_d,F)\right)\right),\label{neutralPhi}
\end{equation} where the functions $N(\mu,F)$, $E_1(\mu,F)$  are
given by (\ref{N(mu,F)}), (\ref{E1(mu,F)}). The factor $1\over2$
accounts for the flavor degeneracy which was included in the
definitions of these functions. The equilibrium
 thermodynamical
potential is obtained by minimizing (\ref{neutralPhi}), for fixed
$\mu$, with respect to $\mu_u,~\mu_d,$ and $F$, under the constraint
(\ref{constraint}).

We find that the behaviors of neutral quark matter and symmetric
quark matter are qualitatively very similar, the pressure of neutral
quark matter being slightly smaller, for a given $\mu$, due to the
charge neutrality constraint.

%%%%%%%%%%%%%%%%%%%%%%%%%%%%%%%%%%%%%%%
\section{Strange quark matter}
In this Section, we consider strange quark matter at densities which
are so high that chiral symmetry has been restored, in the sense
that the masses of quarks $u$, $d$ and $s$ are reduced to the bare
masses.  Moreover, it is assumed for simplicity that the bare masses
vanish. For quarks $s$, this is admittedly a very drastic
assumption. However, we will analyze its consequences and comment at
the end. As in Section 2, we postulate a QCD inspired tensor-tensor
interaction, so that the Lagrangian reads
$${\cal
L}=i\bar\psi\gamma^\mu\partial_\mu\psi-{1\over4}G%[
(\bar\psi\gamma^\mu\gamma^\nu\Lambda_k\psi)(\bar\psi\gamma_\mu\gamma_\nu\Lambda_k\psi)
,$$where $\Lambda_k$ denote the $SU(3)$ flavor Gell-Mann matrices.
This Lagrangian is not chiral invariant. The terms which insure
chiral invariance have been omitted, for simplicity. We assume
polarization along the 3 axis, consider the  mean-field
approximation, and define\begin{equation*} F_k=
iG\langle\bar\psi\gamma_1\gamma_2\Lambda_k\psi\rangle.
\end{equation*}
In the mean field approximation, for $k\neq3,8,$ we have
$F_k=0$, so that the Lagrangian {density} reduces to$${\cal
L}_{MFA}=i(\overline\psi\gamma^\mu\partial_\mu
\psi)-\sum_{k\in\{3,8\}}F_k(\overline\psi
\Sigma_3\Lambda_k\psi)-\sum_{k\in\{3,8\}}{F_k^2\over 2G}.$$ In turn,
this leads to the Dirac equation
$$(-i
\partial_t+i\alpha_j\nabla_j)
\psi-\beta \Sigma_3\sum_{k\in\{3,8\}}F_k \Lambda_k\psi=0.$$ The
operator $\sum_{k\in\{3,8\}}F_k \Lambda_k$ has different values for
quarks $u$, $d$, and $s$. Its eigenvalues %${\cal F}_\tau$
are equal to \begin{equation}{\cal
F}_\tau=\left(F_3+{1\over\sqrt{3}}F_8\right)\delta_{\tau,u}
-\left(F_3-{1\over\sqrt{3}}F_8\right)\delta_{\tau,d}
-{2\over\sqrt{3}}F_8\delta_{\tau,s}\label{Ftau}\end{equation}so that
the Dirac equations reduce to
$$(-i\bfal\cdot{\bfgr}+
{\cal
F}_\tau\beta\Sigma_3)\psi_\tau=\varepsilon_\tau\psi_\tau,\quad\tau\in\{u,d,s\}.$$
We assume that ${\cal F}_u$ is positive and  ${\cal F}_d$, ${\cal
F}_s$ are negative. The single particle energy eigenvalues read
$$\varepsilon_{p,\tau}=\pm\sqrt{\left(|{\cal F}_\tau|\pm\sqrt{p_1^2+p_2^2}\right)^2+p_3^2}~,$$
where the $+$ sign refers to quarks $u$ with momentum $\bf p$ and
spin up and to quarks $d,s$ with the same momentum  and spin down,
while the $-$ sign refers to quarks $u$ with momentum $\bf p$ and
spin down and to quarks $d,s$ with the same momentum  and spin up.
The quark Fermi surfaces are given by
$$\mu^2={\left(|{\cal F}_\tau|\pm\sqrt{p_1^2+p_2^2}\right)^2+p_3^2}~,$$
the $+$ sign corresponding to the Fermi surface of quarks $u$ with
spin up or quarks $d,s$ with spin down, and the $-$ sign to the
Fermi surface of quarks $u$ with spin down or quarks $d,s$ with spin
up.

Several different analytic  expressions, each one with its own range
of validity, are obtained for the thermodynamical potential. One
which is valid when $|{\cal F}_\tau|>\mu$, $\tau\in\{u,d,s\}$,
(needed at high densities), and another for $|{\cal F}_\tau|<\mu$
(needed at low densities). Another one which is valid when the
largest $|{\cal F}_\tau|$ is bigger than $\mu$ (needed at high
densities), and the remaining $|{\cal F}_\tau|$'s are smaller than
$\mu$. Here, $\mu$ is the Fermi energy, which is identified with the
chemical potential. Taking into account that the color degeneracy is
3, in the first case, when full polarization is present, the
$\tau$-quark number reads,
\begin{equation}
N_\tau={{3}V\over(2\pi)^3}\int\d^3{\bf
p}~\theta(\varepsilon_{p,\tau}-\mu)={6V\over(2\pi)^3}\pi^2 |{{\cal
F}_\tau}|\mu^2~, \label{ns}
\end{equation}
and the $\tau$-quark kinetic energy reads,
\begin{equation}
K_\tau={{3}V\over(2\pi)^3}\int\d^3{\bf
p}~\theta(\varepsilon_{p,\tau}-\mu)~\varepsilon_{p,\tau}
={4V\over(2\pi)^3}\pi^2 |{{\cal F}_\tau}|\mu^3 ~. \label{eners}
\end{equation}
\begin{figure}[ht]
\vspace{1.5cm} \centering
\includegraphics[width=0.75\linewidth,angle=0]{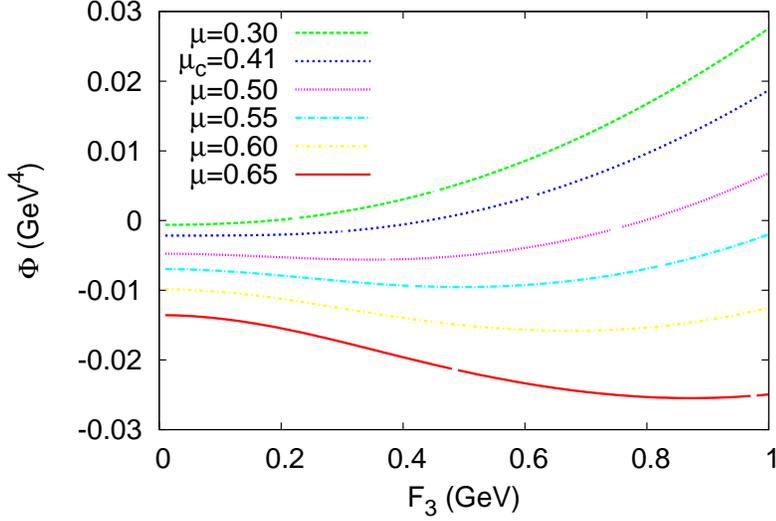} 
\caption{$\Phi$ vs, $F_3$, for strange quark
matter. The curvature at the origin turns upside-down for $\mu$
about $0.410$ GeV.
}
\label{fig6}
\end{figure}
The total quark number and total energy read, respectively
\begin{eqnarray*}&&N={6V\over(2\pi)^3}\pi^2 \mu^2~
(|{{\cal F}_u}|+|{{\cal F}_d}|+|{{\cal F}_s}|),\\&&
E={4V\over(2\pi)^3}\pi^2 \mu^3(|{{\cal F}_u}|+|{{\cal F}_d}|+|{{\cal
F}_s}|)+{V\over2G}(F_3^2+F_8^2),
\end{eqnarray*}
where ${\cal F}_\tau$ is given by (\ref{Ftau}) and the term
${V(F_3^2+F_8^2)/(2G)}$ comes from the standard mean field
approximation.  For full polarization, the thermodynamical potential
becomes
$$\Phi=E-N\mu=-{2V\over(2\pi)^3}\pi^2
(|{\cal F}_u|+|{\cal F}_d|+|{\cal
F}_s|)\mu^3+{V\over2G}(F_3^2+F_8^2)~.$$

As a simplifying ansatz, let us assume that
\begin{equation}\label{approx}F_8={1\over\sqrt{3}}F_3.\end{equation}
This assumption places the quarks $d$ and $s$ on an equal footing.
Then,
\begin{equation}\label{FuFdFs}{\cal F}_u={4\over3}F_3,~{\cal
F}_d={\cal F}_s=-{2\over3}F_3.\end{equation}
\begin{center}
\begin{figure}[!htb]
\subfigure[$G=20$ GeV$^{-2}$]{\includegraphics[scale=0.55]{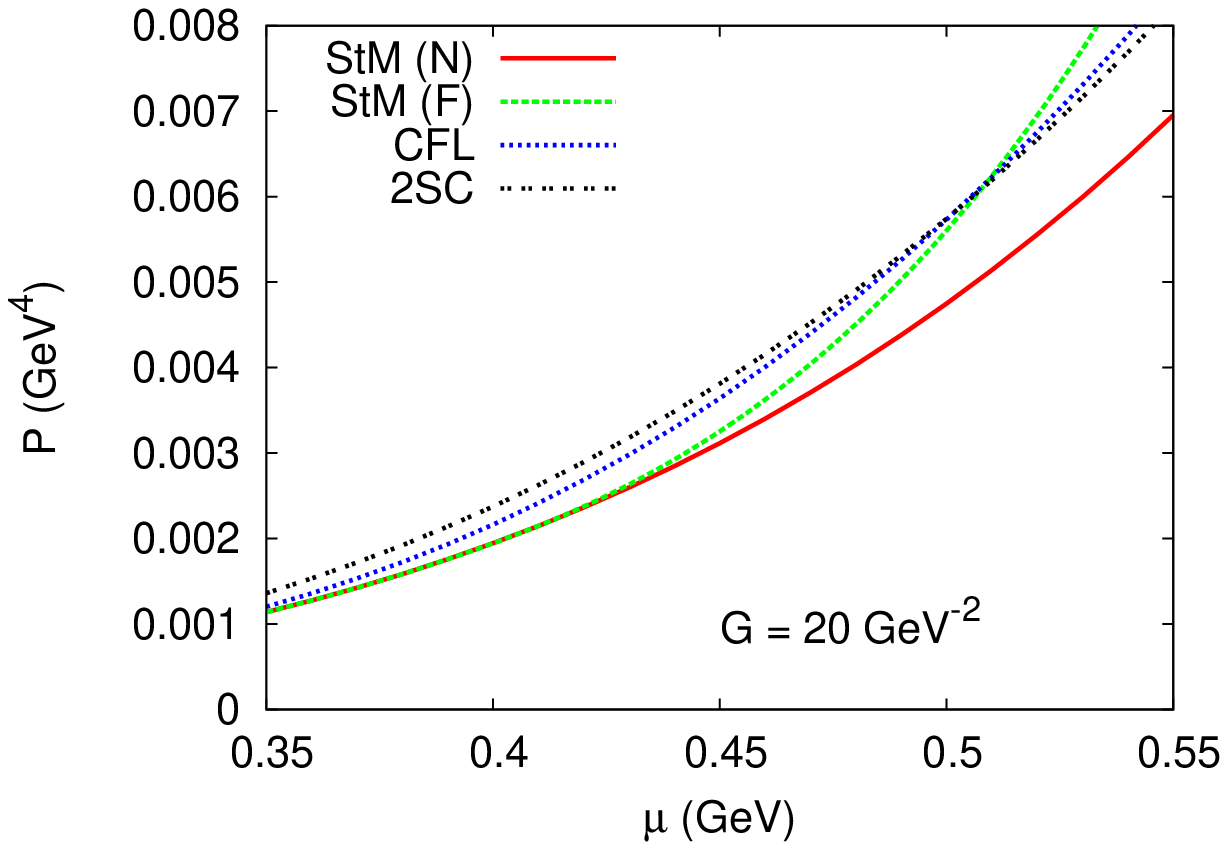}} 
\subfigure[$G=25$ GeV$^{-2}$]{\includegraphics[scale=0.55]{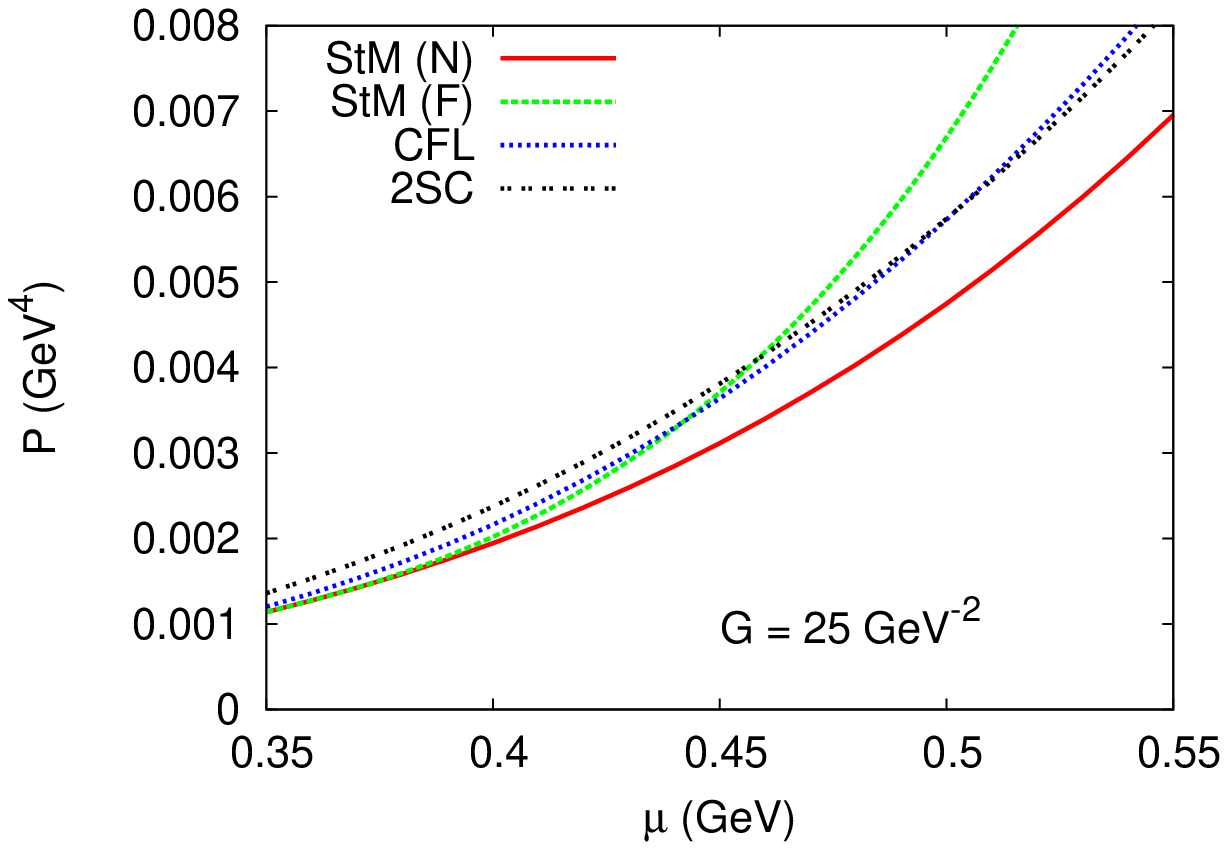}} 
\caption{$P$ vs $\mu$ for strange matter. The
crossing point of the 2SC and spin polarized phases is lower for the
higher $G.$} \label{fig8} \vspace*{-0.2cm}
\end{figure}
\end{center}
Under this ansatz, for ${2\over3}F_3\geq\mu$ (which insures full
polarization, i.e., $|{\cal F}_u|,|{\cal F}_d|,|{\cal F}_s|>\mu$)
the thermodynamical potential equals \begin{equation}\Phi
=-{16V\over3(2\pi)^3}\pi^2|F_3| \mu^3+{2V\over3G}F_3^2
~,\end{equation} implying that
\begin{equation*}F_3=G{\mu^3\over2\pi}~,\end{equation*}so that in
the ferromagnetic phase, which is realized if $\mu$ is above the
critical value $\mu_p$, the thermodynamical potential reads
\begin{equation}\Phi=-{V\over6\pi^2}~{G\mu^6}.\end{equation}
This should be compared with the expression of the thermodynamical
potential  in the normal phase which is realized if $\mu$ is below
the critical value $\mu_c$ and equals
\begin{equation}\Phi=-{3V\over4\pi^2}~{\mu^4}.\end{equation}
Then, in the regime of partial polarization, the thermodynamical
potential of strange quark matter equals
\begin{eqnarray}&&\Phi(\mu,F_3,F_8)={1\over2}(\Phi_1(\mu,{\cal
F}_u)+\Phi_1(\mu,{\cal F}_d) +\Phi_1(\mu,{\cal
F}_s))\nonumber\\&&-{3V\over 8\pi}(({\cal F}_u)^4+({\cal
F}_d)^4+({\cal F}_s)^4)+V{F_3^2+F_8^2\over2G},\end{eqnarray}
\begin{center}
\begin{figure}[!htb]
\subfigure[$G=20$ GeV$^{-2}$] {\includegraphics[scale=0.55]{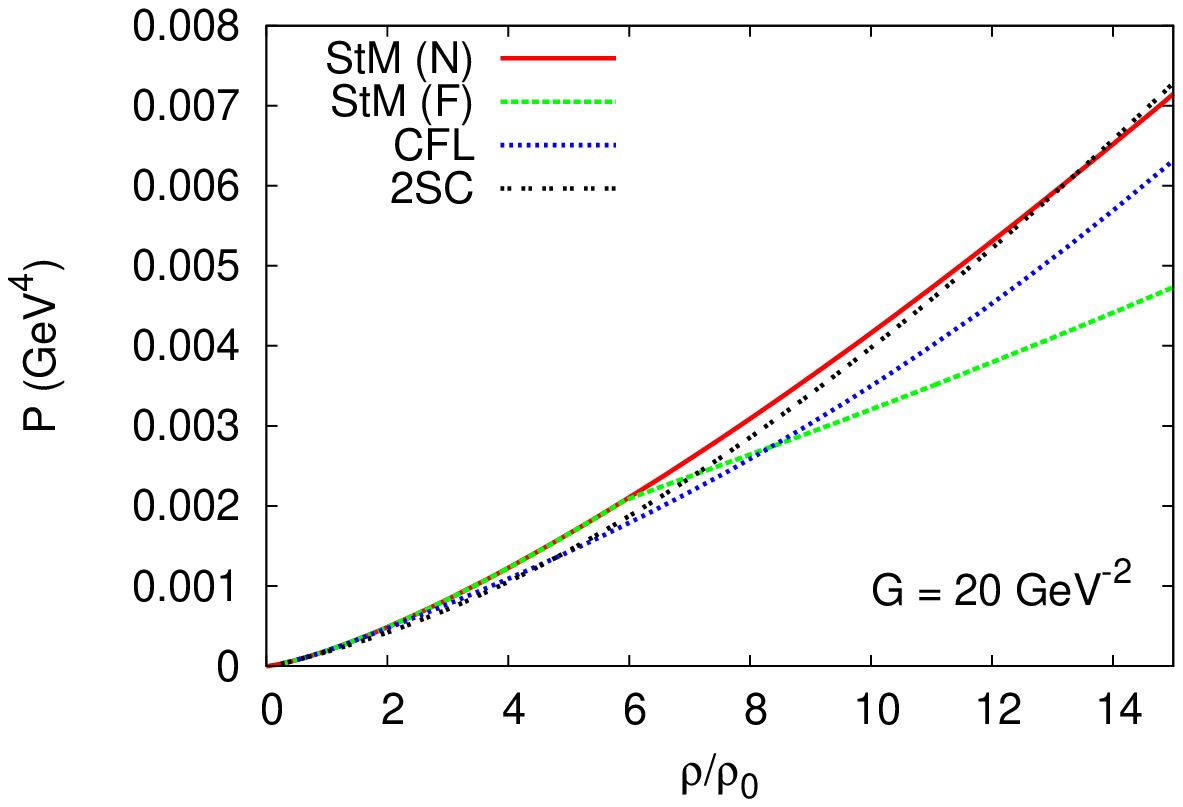}}
\subfigure[$G=25$ GeV$^{-2}$ ]
{\includegraphics[scale=0.55]{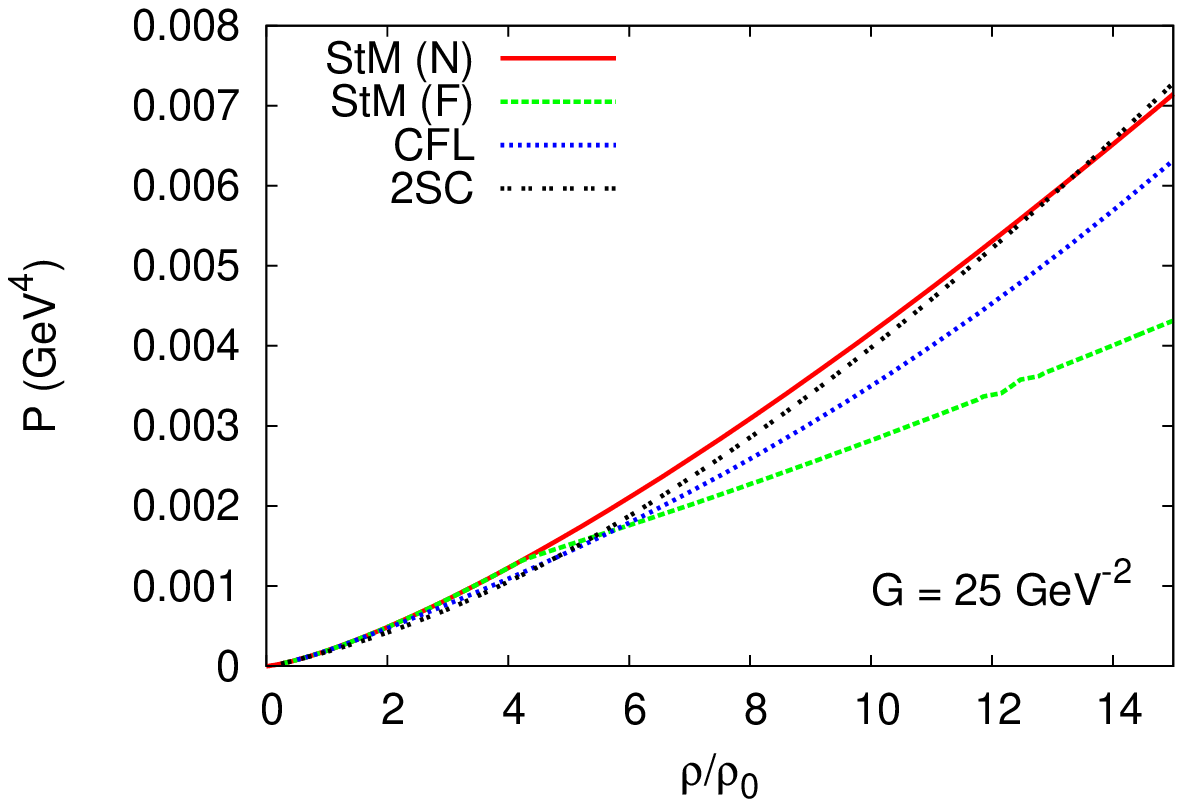}} 
\caption{ The pressure versus the baryonic density, comparing the 2SC 
phase with the spin-polarized phase. }
\label{fig4} \vspace*{-0.2cm}
\end{figure}
\end{center}
where
$\Phi_1(\mu,F)$ is given by (\ref{Phi1}). Under the ansatz
(\ref{approx}), (\ref{FuFdFs}), for $0\leq{4\over3}F_3\leq\mu $ (or
$0\leq{2\over3}F_3\leq{1\over2}\mu$), the thermodynamical potential
reduces to
\begin{eqnarray}&&\Phi(\mu,F_3)={1\over2}\left(\Phi_1\left(\mu,{4\over3}
|F_3|\right)+2\Phi_1\left(\mu,{2\over3}| F_3|\right) \right)-
{8V\over 9\pi} | F_3|^4 +V{2F_3^2
\over3G}\nonumber\\&&.\end{eqnarray}For
${2\over3}F_3<\mu<{4\over3}F_3$ (or
${1\over2}\mu\leq{2\over3}F_3\leq\mu$), the thermodynamical
potential reduces to
\begin{eqnarray} &&\Phi(\mu,F_3)=
\Phi_1\left(\mu,{2\over3}| F_3| \right) -{8V\over 81\pi} |
F_3|^4-{V\over3\pi}F_3\mu^3 +V{2F_3^2 \over3G}.\end{eqnarray}
%%%%%%%%%%%%%%%%%%%%%%%%
\begin{figure}[ht]
\vspace{1.5cm} \centering
\includegraphics[width=0.75\linewidth,angle=0]{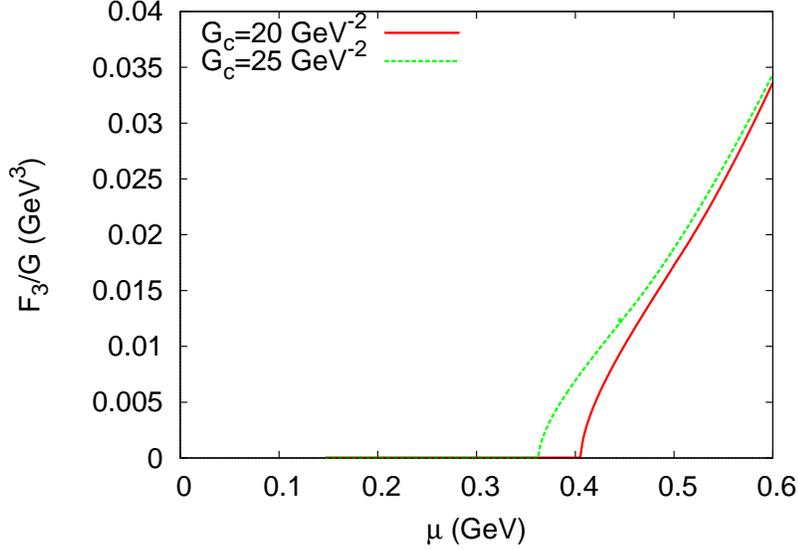}
\caption{The minimizing $F_3$ vs $\mu$, for strange quark matter,
comparing $G=20$ GeV$^{-2}$ with $G=25$ GeV$^{-2}$ The critical
$\mu$ is lower for the higher $G$. } \label{exci7}
\end{figure}
Therefore, we are faced with
three partial expressions. Of course, the equilibrium
thermodynamical potential is obtained by optimizing these partial
expressions in their respective domains of validity. We find that
$\mu_c=\pi/\sqrt{3 G}$ is the critical $\mu$ for the spin
polarization phase transition, and $\mu_{p_2}=\sqrt{3\pi/ G}$ is the
critical $\mu$ for the onset of absolute full polarization. The
expression of the critical chemical potential  $\mu_{p_1}$  for the
onset of full polarization of quarks $u$ (but not of quarks $d,s$)
is not given because it is cumbersome.

Color superconductivity in quark matter has been investigated in
\cite{bohr2}, where a  color-flavor symmetric version of the
so-called 2 flavor super-conducting (2SC) phase has been proposed
and its transition to so-called color-flavor locked phase (CFL) was
analyzed.  In \cite{bohr2}, color superconductivity has been
described by a mean-field thermodynamical potential operator (i.e.,
the Hamiltonian constrained by the quark number conservation) of the
form,
\begin{eqnarray}\hat
K_{MFA}&=&\sum_{iap}\epsilon_pc^\dagger_{iap}c_{iap}\nonumber\\
&+&\sum_{ijkabcp}( \Delta^*_{kc} ~{c_{jbp}~c_{ia\overline
p}}~\epsilon_{ijk}\epsilon_{abc}\eta_p+ \Delta_{kc}
~c^\dagger_{iap}~c^\dagger_{jb\overline
p}~\epsilon_{ijk}\epsilon_{abc}\eta_p)+{V\over
2G_c}\sum_{jc}\Delta_{jc}\Delta_{jc}^*,\nonumber\\\Delta_{kc}&=&{G_c\over
V}\sum_{ijabp}\langle {c_{jbp}~c_{ia\overline
p}}\rangle~\epsilon_{ijk}\epsilon_{abc}\eta_p,\label{KMFA}
\end{eqnarray}
where $\epsilon_p=\sqrt{p^2+M^2}-\mu$, $M$ being the current mass.
In the  2SC phase  presented in \cite{bohr2}, the gaps $\Delta_{jc}$
are real and independent of the color-flavor indices, that is,
$\Delta_{jc}=\Delta,~j\in\{u,d,s\},c\in\{r,g,b\}$, so that the two
specific flavors or the two specific colors which are paired are not
pure colors or flavors, but are related to pure colors and flavors
through a certain rotation in color-flavor space, which has the
great advantage of insuring automatically color-flavor symmetry. In
the CFL phase, the indices $j,c$ are coupled in a prescribed way,
for instance, $(u$ with $r),(d$ with $g),(s$ with  $b)$ (cf.
\cite{bohr2} for further details). For the model considered in
\cite{bohr2}, the 2SC phase is stable for $\mu<0.505$ GeV, while the
CFL phase is stable for $\mu>0.505$ GeV, for $G_c=5$GeV$^{-2}$. In
that model, the quarks have all the same helicity (either positive
or negative). If we wish to take into account both helicities,  the
gaps reported in \cite{bohr2} are valid and the pressure becomes
twice the pressure reported there, under the condition that half the
value of $G_c$ is used, that is, for $G_c=2.5$GeV$^{-2}$.

It is of interest to investigate the interplay between color
superconductivity and spin polarization as is here described. The
transition between the 2SC  phase and the ferromagnetic phase is
depicted in Fig. \ref{fig8},
%(in more detail in fig. \ref{fig8}),
where the pressure $P$ is plotted vs the chemical potential $\mu$
for the normal phase, for the ferromagnetic phase, for the 2SC phase
and for the CFL phase.  The curves corresponding to the spin
polarized F phase, to the 2CS phase and to the CFL phase seem to
cross at the same point. This is by pure accident. If instead of
$G=20$ GeV$^{-2}$ a slightly larger value is considered, the F curve
will cross the 2CS and the CFL curves before these curves cross each
other.  This is clearly seen in fig \ref{fig8} (b), where the spin
polarized phase for $G=25$ GeV$^{-2}$ is plotted. In our model, the
CFL phase seems to be completely hidden by the ferromagnetic phase,
but, if a slightly lower $G $ is used, the CFL phase will be
realized for a short while. Spin polarization and color
superconductivity has recently been considered in \cite{nakano}. The
relevance of the interplay between the CFL phase and the 2SC phase
for the stability of hybrid stars is discussed in \cite{bub}.

 Suppose we have an unperturbed Lagrangian ${\cal L}_0$ and two
types of perturbations,   ${\cal L}_I$ and ${\cal L}_{II}$, so that
${\cal L}_0+{\cal L}_{I}$ describes the phase $I$ associated with
some condensate and  ${\cal L}_0+{\cal L}_{II}$ describes the phase
$II$ associated with another condensate. The interplay between the
two phases is described by the Lagrangian ${\cal L}_0+{\cal
L}_{I}+{\cal L}_{II}$. However, in order to treat approximately this
interplay we may apply the Gibbs criterium to the first order
transition between the phases described by the truncated
Lagrangians. Clearly, it would be more correct to use the full
Lagrangian, following \cite{kitazawwa}, and then the first order
transition may be replaced by a second order one. We have used the
approximate method to describe the interplay between the 2SC phase
and the spin polarized phase.

\section{Conclusions}
Spin polarization of a high density hadronic fluid of quarks has
been studied. We describe a
%ferromagnetic
spin polarized phase in hadronic matter arising from a four Fermion
interaction which may be regarded as the relativistic analogue of
the interaction considered in \cite{navarro,skyrme,vidana} and is
related to the interaction of the standard NJL model through a Fierz
transformation. A {\it no sea} approximation has been used.  Flavor
$SU(2)$ and flavor $SU(3)$ quark matter are investigated.
%In flavor $SU(2)$ quark matter, a
A second order phase transition  at densities about 5 times the
normal nuclear matter density is predicted and a magnetization of
the order of $10^{16}$ gauss is expected. Numerical results for
$G=20$ GeV$^{-2}$, which is arbitrary but of the order of the
coupling constant in the NJL model, predict a phase transition at
$5-8$ times the saturation density $\rho_0$, which is probably too
high. Increasing the coupling to $G=30$ GeV$^{-2}$ would decrease
the transition density to $3\, \rho_0$.   It is also found that in
flavor $SU(3)$ quark matter, a first order transition from the 2SC
phase \cite{bohr2} to the spin polarized phase arises. The spin
polarized phase is stable for $\mu>0.51$ GeV. The 2SC phase is
stable for $\mu<0.51$ GeV.
%{\color{magenta}What is the density? $\rho/\rho_0$?}
In our model, the CFL phase \cite{bohr2} is
completely overshadowed by the spin polarized phase.

We have taken $G=20$ GeV$^{-2}.$ The critical chemical potential for
symmetric quark matter and for strange matter turns out to be
$\mu_c=$0.41 GeV. If lower $G$ values are considered, $\mu_c$
increases and in that case the transition from the 2SC to the CFL
phase occurs before the transition to the spin polarized phase, so
that the CFL phase is realized. If higher $G$ values are considered,
$\mu_c$ decreases, so that the transition to the spin polarized
phase occurs before the transition from the 2SC phase to the CFL
phase, which is then completely overshadowed. Due to the vacuum
polarization, it is still meaningful to consider slightly higher $G$
values. Indeed, the main effect of taking the vacuum polarization
into account, is the appearance, in the expression of the
thermodynamical potential, of a new term of the form $-\kappa
\Lambda^2 F^2$, where $\kappa>0$ is a certain numerical factor and
$\Lambda$ is the regularization cutoff. But this will only
renormalize the coupling constant $G$, so that we may disregard the
vacuum polarization provided we replace the unrenormalized $G_0$ by
a new $G$ such that $-\kappa\Lambda^2+1/(2G_0)=1/(2G)$, implying
that $G_0<G$. This may
%be the reason why the here used coupling constant turns out to be somewhat too
justify the use of a large $G$. For simplicity, we have assumed that
the strange quark current mass vanishes. This is an acceptable
assumption for $\mu_c=$0.41 GeV.

This has been an exploratory work where the possibility of a phase
transition to a magnetized phase in quark matter was investigated.
The application to a physical system like a compact star may require
the careful consideration of the vacuum contribution and of finite
quark current masses.

\section*{Acknowledgements}
The present research was partially supported by the projects
FCOMP-01-0124-FEDER-008393 with FCT reference
%CERN/FP/109316/2009, PTDC ./FIS /64707/2006 and
PTDC/FIS/113292/2009.

%XXXXXXXXXXXXXXXXXXX


\begin{thebibliography}{www}
\bibitem{haensel} P. Haensel, A.Y. Potekhin and D.G. Yakovlev,
{\it Neutron  Stars 1} Springer, ISBN 10-0-387-33543-9 (2007).
\bibitem{g}
R. C. Duncan and C. Thompson, The Astrophys. J. 392 (1992) L9; C.
Thompson and R. C. Duncan, MNRAS 275(1995) 255 ; V. V. Usov, Nature
357(1992) 472 ; B. Paczynski, Acta Astron. 42 (1992) 145 .
%R. C. Duncan and C. Thompson, {\it Magnetars} in High velocity
%neutron stars and gamma-ray bursts, eds. R.E. Rotschild and R.E.
%Lingenfelter, AIP Conf. Proc. 366, Am. Imst. Phys. New York, 1966.
\bibitem{goegues}
E. G\"o\u g\"us et al., The Astrophysical J. 718 (2010) 331.
\bibitem{bohr} H. Bohr, C. Provid\^encia and J. da Provid\^encia,
 Braz. J. Phys. 42 (2012) 68.
\bibitem{njl} Y. Nambu and G. Jona-Lasinio, Phys Rev. {\bf 122} (1961) 345 .
\bibitem{klevansky} S.P. Klevanky, Rev. Mod. Phys. 64 (1992) 649.
\bibitem{hatsuda} T.Hatsuda and T.Kunihiro, Phys. Rep. 247 (1994) 221.
\bibitem{buballa} M. Buballa, Phys. Rep. 407 (2005) 205.
\bibitem{broniowski} M. Kutschera, W. Broniowski and A. Kotlorz, Nucl.
Phys. A516 (1990) 566.
%\bibitem{dautry} F. Dautry and E. M. Nyman, Nucl. Phys. A319 (1979)
%323.
%\bibitem{takahashi} K. Takahashi, J. Phys. G: Nucl. Part. Phys. 32
%(2006) 1131-1141; 32 (2006) 2347.
%\bibitem{broderick}A. Broderick , M. Prakash , and J. M. Lattimer,
%The Astrophysical Journal, 537(2000)351-367.
%\bibitem{chakrabarty} S. Gosh, S. Mandal and S. Chakrabarty, Phys.
%Rev. C75 (2007) 015805.
%\bibitem{menezes}D. P. Menezes, M. Benghi Pinto, S. S. Avancini, A.
%P\'erez Mart\'\i nez, and C. Provid\^encia, Phys. Rev. C 79 (2009)
%035807.
%\bibitem{ebert}D. Ebert and K. G. Klimenko,
%Nucl. Phys. A728 (2003) 203.;  D. Ebert, K. G. Klimenko, M. A.
%Vdovichenko, and A. S. Vshivtsev, Phys. Rev. D 61, 025005 (1999).
\bibitem{navarro}
A. Vidaurre, J. Navarro and J. Bernab\'eu, Astron. Astrophys. 135
(1984) 361.
\bibitem{skyrme} J. Margueron, J. Navarro, and Nguyen Van Giai, Phys. Rev. C 66 (2002) 014303.
\bibitem{vidana} A. Rios, A. Polls and I. Vida\~na, Phys. Rev. C71
(2005) 055802.
\bibitem{isaac02} I. Vida\~na, A. Polls, and A. Ramos, Phys. Rev. C 65 (2002) 035804.
\bibitem{tatsumi} T. Tatsumi, Phys. Lett B489 (2000) 132.
\bibitem{iwazaki}A. Iwazaki, Phys. Rev. D 72  (2005) 114003.
\bibitem{ebert}
D. Ebert, V. Zhukovsky, O.V. Tarasov,
%"Competition of color
%ferromagnetic and superconductive states in a quark-gluon system",
%Phys Rev D, 72:9 (2005), 096007
%"Competition of color ferromagnetic and superconductive
%states in a quark-gluon system",
Phys. Rev. D 72 (2005) 096007.
%>
%> Tatsumi, Phys.Lett. B489 (2000) 280-286
%> Ebert et al., Phys.Rev. D72 (2005) 096007
%> Iwazaki et al., Int.J.Mod.Phys. A22 (2007) 721-730
%> Modarres et al.,  Physica A387 (2008) 2761-2776
\bibitem{bohr2} H. Bohr, P. K. Panda, C. Provid\^encia and J. da
Provid\^encia, Braz. J. Phys. 42 (2012) 59.
\bibitem{nakano} E. Nakano, T. Maruyama and T. Tatsumi, Phys. Rev. D
68 (2003) 1051001.
\bibitem{bub} M. Buballa, F. Neumann, M. Oertel and I. Shovkovy
   Phys. Lett. {\bf B595} (2004) 36.
\bibitem{kitazawwa} M. Kitazawwa, T. Koide, T. Kunihiro and Y.
Nemoto, Prog. Theor. Phys. 108 (2002) 929.
\end{thebibliography}
\end{document}